\documentclass{iau}

\usepackage{amsmath}
\usepackage{graphicx}
\usepackage{multirow}
\usepackage{bm}
\usepackage{natbib}
\usepackage{comment}
\newcommand{\Fig}[1]{Figure~\ref{#1}}

\begin{document}

\lefttitle{Cambridge Author}
\righttitle{Proceedings of the International Astronomical Union: \LaTeX\ Guidelines for~authors}

\jnlPage{1}{7}
\jnlDoiYr{2021}
\doival{10.1017/xxxxx}

\aopheadtitle{Proceedings IAU Symposium}
\editors{A. V. Getling \&  L. L. Kitchatinov, eds.}

\title{The role of nonlinear toroidal flux loss due to flux
emergence in the long-term evolution of the solar cycle}

\author{Akash Biswas}
\affiliation{Department of Physics, Indian Institute of Technology (Banaras Hindu University), Varanasi 221005, India}

\begin{abstract}
A striking feature of the solar cycle is that at the beginning, sunspots appear around mid-latitudes, and over time the latitudes of emergences migrate towards the equator. The maximum level of activity varies from cycle to cycle. For strong cycles, the activity begins early and at higher latitudes with wider sunspot distributions than for weak cycles. The activity and the width of sunspot belts increase rapidly and begin to decline when the belts are still at high latitudes. However, in the late stages of the cycles, the level of activity, and properties of the butterfly wings all have the same statistical properties independent of the peak strength of the cycles. We have modelled these features using 
Babcock–Leighton type dynamo model and shown that the toroidal flux loss from the solar interior due to magnetic buoyancy is an essential nonlinearity that leads to all the cycles decline in the same way.
\end{abstract}


\maketitle

\section{Introduction}

The magnetic activity of the Sun, as generally measured by the presence of sunspots (or more generally bipolar magnetic regions) on the solar surface evolves in a cyclic fashion with a period of about 11 years, commonly known as the `Solar Cycle' \citep{Hat15}. The long term study of the solar magnetic activity reveals that the 11 year solar cycles possess significant variations in their characteristics which makes the otherwise similar looking cycles unique and widely different from each other \citep{Karak23,Uso13, BKUW23}.

The strong cycles rise faster and take less time to reach peak, whereas the weaker ones rise slowly and take more time to attain their peak strength. This phenomena is known as the 
Waldmeier effect  \citep{W35, KC11} which is also detected in the magnetic cycles of other solar-type stars \citep{garg19}. 
A more detailed look at the rising phases of the solar cycles reveals that the stronger cycles start early, and shows sunspot activities on higher latitudes, with wider latitudinal distribution in the initial phase, on the other hand, the weak cycles exhibit a delayed start of their activity and from lower latitude regions with narrower latitudinal bands of activity. Hence the rising phases of the cycles are quite different from each other depending on their eventual strength.

In 1955, Waldmeier reported about an even more intriguing aspect of the evolution of the solar cycles is that, during their decline phases of the cycles, they evolve with similar statistical properties irrespective of their strength \citep{W55}. 
Later, \cite{CS16} analysed the century-scale 
sunspot observations and found out that when the cycles are represented in terms of the average latitude and width of their annual activity belt and the annual strength of their activity, the rising phases of the cycles show different trajectories of evolution however, soon after the cycles reach their peak, they evolve with a common trajectory, i.e. all the cycles decay in the same way with similar statistical properties irrespective of their peak strength.

The phenomena of solar cycles evolving in different manners through their rising phases, but possessing similar properties while declining indicates towards the presence of a nonlinearity in solar dynamo process. In this work, we perform numerical simulations of the solar dynamo to explore the role of this nonlinearity behind the aforementioned traits of the solar cycle. 

\section{Model description:}
The underlying reason behind the sustained cyclic variation of strength of the solar magnetic fields is believed to be the dynamo mechanism working in the solar convection zone. In recent times, the Babcock-Leighton \citep{Ba61, Leighton69} type solar dynamo models have made significant progress in explaining various observed features of magnetic activity of the Sun and Sun like stars  \citep{Char20, Kar10, KM18, KMB18, VKK21, VKK23} and have been extensively used for forecasting solar cycle strength \citep{CCJ07,Bhowmik+Nandy, Kumar21, KBK22, BKK23}. 

In this work, we are utilizing a Babcock-Leighton type, axisymmetric, kinematic solar dynamo code `SURYA' \citep{NC02, CNC04, KC12} which solves the following two coupled equations to capture the evolution of the poloidal and toroidal component of the solar magnetic field:

\begin{equation}
\frac{\partial A}{\partial t} + \frac{1}{s}({\bm v_p} \cdot {\bf \nabla})(s A) = \eta_p\left(\nabla^2 -\frac{1}{s^2}\right)A + \alpha B,
\label{eq:pol}
\end{equation} 

\begin{equation}
\frac{\partial B}{\partial t} + \frac{1}{r}\left[\frac{\partial (rv_rB)}{\partial r}+ \frac{\partial (v_\theta B)}{\partial \theta}  \right] = \eta_t\left(\nabla^2 - \frac{1}{s^2}\right)B + s({\bm B_p} \cdot {\bf \nabla})\Omega + \frac{1}{r}\frac{d\eta _t}{dr}\frac{\partial (rB)}{\partial r},
\label{eq:tor}
\end{equation}

Where, $s = r \sin \theta$, $A$ is the magnetic potential of the poloidal magnetic field, $B$ is the toroidal magnetic field. A detailed discussion about the model parameters for this particular work can be found in \cite{BKC22}. 

Here we describe the source of nonlinearity in this dynamo model which is related to the process of buoyant rise of the toroidal flux, i.e. the formation of sunspots. In every certain time  interval it is checked whether the strength of the toroidal magnetic field in the convection zone has crossed a critical value $B_c$. If at any grid point the field is found to have crossed $B_c$, it is assumed that flux tube will buoyantly emerge to the surface. The value  of the magnetic field is locally reduced by half and the other half is added to the surface. This is how the formation of sunspot due to the buoyant rise of the toroidal flux tubes are captured in the model in a simple way to ensure that with each sunspot eruption, a part of the toroidal flux is lost from the solar convection zone. The value of the critical field is taken as $B_c = 0.8 \times 10^4 ~ G$ to keep the surface radial field strength within the observed range \citep{Mord22, GBKKK23}, and to keep the total amount of toroidal flux loss consistent with observational estimates \citep{CS20}. It is worth mentioning that, just by constraining the model parameters from observations, we find that the value of the critical field $B_c$ is similar to the equipartiiton field strength at the base of solar convection zone, as calculated from the mixing length approximation. 

\section{Results and Discussions}
\begin{figure}
\centering
\includegraphics[scale=0.26]{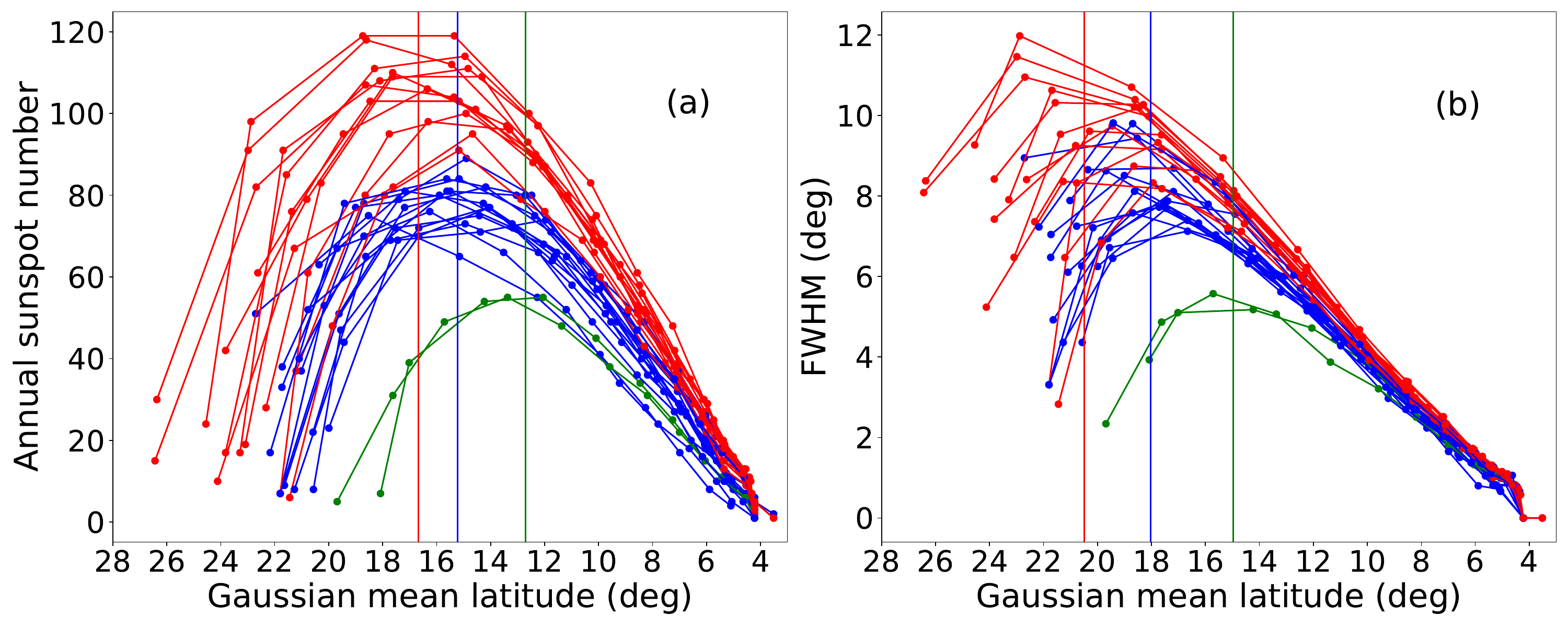}
\caption{ (a) The model sunspot number for each year plotted as the function of the mean latitude of the annual sunspot distribution. Each curve correspond to one cycle. (b) Same  as (a) but with the width of the annual latitudinal distribution. The cycles begin from the left (at higher latitudes) and progress towards the right with time. The vertical lines correspond to the average latitude where the strong (red), moderate (blue), and weak (green) cycles reach their peak.
}
\label{fig:CS}
\end{figure}

Following the model setup described in the previous section, we perform simulation spanning over 40 cycles. The (pseudo) sunspots for each of the cycles from the simulation are tracked by the latitude of their emergence. The annual latitudinal distribution of the sunspots approximately follow a Gaussian distribution similar to what is seen in observation \citep{CS16}. From these distributions, the central latitude of the activity belts and the width of their distribution are measured. Here in  the \Fig{fig:CS}, the trajectories of the cycles are shown as the sunspot activity belts migrate from high latitude regions towards the equator with the progress of the cycle. It can be clearly seen that during the beginning phases of the cycles, when the sunspot activities are centred on higher latitude regions, the cycles evolve very differently from each other, the strong cycles start early at higher latitudes, with wider activity belts, they rise rapidly and attain their peak at higher latitudes, on the other hand, the weaker cycles, start lately, from somewhat lower latitude regions, rise slowly and eventually attain their maxima at even lower latitudes. On the other hand, during the decline phases of the cycles, the trajectories merge quickly after obtaining the peaks and the cycles evolve with similar properties. Hence, the observed features of the latitudinal distribution of the solar magnetic activity throughout the phases of solar cycle as described earlier is very well reproduced in this model.

Now, let us discuss the role of the nonlinear flux loss in the evolution of solar cycles with different strengths. In a strong cycle, the poloidal field strength during the minima of the cycle is strong, as a result the rate of production of the toroidal field due to the action of differential rotation is high, hence the toroidal flux emergence condition is fulfilled very early in the cycle in a wider latitudinal region. As a result, the sunspot activity for these cycles begin very early with a wider latitudinal distribution and at higher latitudes. Due to the high production rate of the toroidal field, the emergence rate of the sunspots are also high here, however, with each sunpsot emergence, a significant portion of the toroidal flux is lost from the solar interior. As a result, the toroidal field strength, attains saturation very quickly at high latitudes, when the subsurface toroidal field strength becomes comparable to $B_c$, marking the maxima of the solar cycle. During the declining phase of the cycles, any further production of the toroidal flux is taken care of by the further loss of flux due to emergences as the activity belts get advected towards the equator by the meridional circulation. 

In the case of a weak cycle, the poloidal field during the minima is weak and as a result, the toroidal flux is generated at a low pace. Hence, it takes more time for the toroidal field to meet the threshold for flux emergence. By the time the toroidal field becomes strong enough to produce sunspots, the merdiional circulation transports the activity belt to lower latitudes, this explains why the weak cycles exhibit a delayed start and begin their journey from a low latitude. Also, due to the low rate of toroidal field production, lead to the flux eruption condition getting satisfied only at a narrow latitudinal regions and the activity of the cycle rise slower than the stronger cycles. The weak cycles lose less amount of toroidal flux due to their low rise rate, hence the toroidal field takes more time to be saturated and in the meanwhile, the meridional circulation advects the activity belts to further lower latitudes making the weaker cycles reach their peak at lower latitudes in comparison to the stronger ones. However, during the decline phase, the weaker cycles evolve exactly the same way as the stronger ones, the value of the toroidal field stays comparable to the critical value $B_c$ and any further build-up of the toroidal flux is taken care of by the further flux lost due to emergencies.

\section{Conclusion}
In conclusion, we find that a strong nonlinear mechanism of toroidal flux loss due to flux emergence is essential to reproduce the observed features of the latitudinal distributions of solar cycle activity. This mechanism significantly enhances the amount and rate of flux lost during the rising phases of the strong cycles in comparison to weak cycles ensuring that the morphology of the toroidal magnetic field inside the Sun's convection zone becomes similar for all cycles during their decline phases, hence all the cycles decline in the same way having similar statistical properties irrespective of their peak strength. 
By comparing the model outputs with observations, it has been found that, the critical field strength ($B_c$) for the threshold of flux emergence is similar to the equipartition field strength at solar convection zone.
Our study also provides an explanation of the so-called latitudinal quenching---strong cycles produce BMR at high latitudes \citep{MKB17}, which in turn generates a weak polar field and stabilizes the solar dynamo \citep{J20, Kar20}

\section{Acknowledgements}
I thank the International Astronomical Union for the award of the generous travel grant to attend the in-person IAU Symposium 365, in Yerevan Armenia. 
The financial support from the University Grants Commission, Govt. of India is gratefully acknowledged.

\bibliographystyle{iaulike}
\bibliography{iautalk} 

\begin{thebibliography}{}

\bibitem[{Babcock}, 1961]{Ba61}
{Babcock}, H.~W. 1961, {The Topology of the Sun's Magnetic Field and the
  22-YEAR Cycle.}
\newblock {\em \apj}, 133, 572.

\bibitem[{Bhowmik} and {Nandy}, 2018]{Bhowmik+Nandy}
{Bhowmik}, P. \& {Nandy}, D. 2018, {Prediction of the strength and timing of
  sunspot cycle 25 reveal decadal-scale space environmental conditions}.
\newblock {\em Nature Communications}, 9, 5209.

\bibitem[{Biswas} et~al., 2022]{BKC22}
{Biswas}, A., {Karak}, B.~B., \& {Cameron}, R. 2022, {Toroidal Flux Loss due to
  Flux Emergence Explains why Solar Cycles Rise Differently but Decay in a
  Similar Way}.
\newblock {\em \prl}, 129(24), 241102.

\bibitem[{Biswas} et~al., 2023a]{BKK23}
{Biswas}, A., {Karak}, B.~B., \& {Kumar}, P. 2023,a {Exploring the reliability
  of polar field rise rate as a precursor for an early prediction of solar
  cycle}.
\newblock {\em \mnras}, 526a(3), 3994--4003.

\bibitem[{Biswas} et~al., 2023b]{BKUW23}
{Biswas}, A., {Karak}, B.~B., {Usoskin}, I., \& {Weisshaar}, E. 2023,b
  {Long-Term Modulation of Solar Cycles}.
\newblock {\em \ssr}, 219b(3), 19.

\bibitem[{Cameron} and {Sch{\"u}ssler}, 2016]{CS16}
{Cameron}, R.~H. \& {Sch{\"u}ssler}, M. 2016, {The turbulent diffusion of
  toroidal magnetic flux as inferred from properties of the sunspot butterfly
  diagram}.
\newblock {\em \aap}, 591, A46.

\bibitem[{Cameron} and {Sch{\"u}ssler}, 2020]{CS20}
{Cameron}, R.~H. \& {Sch{\"u}ssler}, M. 2020, {Loss of toroidal magnetic flux
  by emergence of bipolar magnetic regions}.
\newblock {\em \aap}, 636, A7.

\bibitem[{Charbonneau}, 2020]{Char20}
{Charbonneau}, P. 2020, {Dynamo models of the solar cycle}.
\newblock {\em Living Reviews in Solar Physics}, 17(1), 4.

\bibitem[{Chatterjee} et~al., 2004]{CNC04}
{Chatterjee}, P., {Nandy}, D., \& {Choudhuri}, A.~R. 2004, {Full-sphere
  simulations of a circulation-dominated solar dynamo: Exploring the parity
  issue}.
\newblock {\em \aap}, 427, 1019--1030.

\bibitem[{Choudhuri} et~al., 2007]{CCJ07}
{Choudhuri}, A.~R., {Chatterjee}, P., \& {Jiang}, J. 2007, {Predicting Solar
  Cycle 24 With a Solar Dynamo Model}.
\newblock {\em Physical Review Letters}, 98(13), 131103.

\bibitem[{Garg} et~al., 2019]{garg19}
{Garg}, S., {Karak}, B.~B., {Egeland}, R., {Soon}, W., \& {Baliunas}, S. 2019,
  {Waldmeier Effect in Stellar Cycles}.
\newblock {\em \apj}, 886(2), 132.

\bibitem[{Golubeva} et~al., 2023]{GBKKK23}
{Golubeva}, E.~M., {Biswas}, A., {Khlystova}, A.~I., {Kumar}, P., \& {Karak},
  B.~B. 2023, {Probing the variations in the timing of the Sun's polar magnetic
  field reversals through observations and surface flux transport simulations}.
\newblock {\em \mnras}, 525(2), 1758--1768.

\bibitem[{Hathaway}, 2015]{Hat15}
{Hathaway}, D.~H. 2015, {The Solar Cycle}.
\newblock {\em Living Reviews in Solar Physics}, 12(1), 4.

\bibitem[{Jiang}, 2020]{J20}
{Jiang}, J. 2020, {Nonlinear Mechanisms that Regulate the Solar Cycle
  Amplitude}.
\newblock {\em \apj}, 900(1), 19.

\bibitem[{Karak}, 2010]{Kar10}
{Karak}, B.~B. 2010, {Importance of Meridional Circulation in Flux Transport
  Dynamo: The Possibility of a Maunder-like Grand Minimum}.
\newblock {\em \apj}, 724, 1021--1029.

\bibitem[{Karak}, 2020]{Kar20}
{Karak}, B.~B. 2020, {Dynamo Saturation through the Latitudinal Variation of
  Bipolar Magnetic Regions in the Sun}.
\newblock {\em \apjl}, 901(2), L35.

\bibitem[{Karak}, 2023]{Karak23}
{Karak}, B.~B. 2023, {Models for the long-term variations of solar activity}.
\newblock {\em Living Reviews in Solar Physics}, 20(1), 3.

\bibitem[{Karak} and {Choudhuri}, 2011]{KC11}
{Karak}, B.~B. \& {Choudhuri}, A.~R. 2011, {The Waldmeier effect and the flux
  transport solar dynamo}.
\newblock {\em \mnras}, 410, 1503--1512.

\bibitem[{Karak} and {Choudhuri}, 2012]{KC12}
{Karak}, B.~B. \& {Choudhuri}, A.~R. 2012, {Quenching of Meridional Circulation
  in Flux Transport Dynamo Models}.
\newblock {\em \solphys}, 278, 137--148.

\bibitem[{Karak} et~al., 2018]{KMB18}
{Karak}, B.~B., {Mandal}, S., \& {Banerjee}, D. 2018, {Double Peaks of the
  Solar Cycle: An Explanation from a Dynamo Model}.
\newblock {\em \apj}, 866(1), 17.

\bibitem[{Karak} and {Miesch}, 2018]{KM18}
{Karak}, B.~B. \& {Miesch}, M. 2018, {Recovery from Maunder-like Grand Minima
  in a Babcock--Leighton Solar Dynamo Model}.
\newblock {\em \apjl}, 860, L26.

\bibitem[{Kumar} et~al., 2022]{KBK22}
{Kumar}, P., {Biswas}, A., \& {Karak}, B.~B. 2022, {Physical link of the polar
  field buildup with the Waldmeier effect broadens the scope of early solar
  cycle prediction: Cycle 25 is likely to be slightly stronger than Cycle 24}.
\newblock {\em \mnras}, 513(1), L112--L116.

\bibitem[{Kumar} et~al., 2021]{Kumar21}
{Kumar}, P., {Nagy}, M., {Lemerle}, A., {Karak}, B.~B., \& {Petrovay}, K. 2021,
  {The Polar Precursor Method for Solar Cycle Prediction: Comparison of
  Predictors and Their Temporal Range}.
\newblock {\em \apj}, 909(1), 87.

\bibitem[{Leighton}, 1969]{Leighton69}
{Leighton}, R.~B. 1969, {A Magneto-Kinematic Model of the Solar Cycle}.
\newblock {\em \apj}, 156, 1.

\bibitem[{Mandal} et~al., 2017]{MKB17}
{Mandal}, S., {Karak}, B.~B., \& {Banerjee}, D. 2017, {Latitude Distribution of
  Sunspots: Analysis Using Sunspot Data and a Dynamo Model}.
\newblock {\em \apj}, 851, 70.

\bibitem[{Mordvinov} et~al., 2022]{Mord22}
{Mordvinov}, A.~V., {Karak}, B.~B., {Banerjee}, D., {Golubeva}, E.~M.,
  {Khlystova}, A.~I., {Zhukova}, A.~V., \& {Kumar}, P. 2022, {Evolution of the
  Sun's activity and the poleward transport of remnant magnetic flux in Cycles
  21-24}.
\newblock {\em \mnras}, 510(1), 1331--1339.

\bibitem[{Nandy} and {Choudhuri}, 2002]{NC02}
{Nandy}, D. \& {Choudhuri}, A.~R. 2002, {Explaining the Latitudinal
  Distribution of Sunspots with Deep Meridional Flow}.
\newblock {\em Science}, 296(5573), 1671--1673.

\bibitem[{Usoskin}, 2013]{Uso13}
{Usoskin}, I.~G. 2013, {A History of Solar Activity over Millennia}.
\newblock {\em Living Reviews in Solar Physics}, 10, 1.

\bibitem[{Vashishth} et~al., 2021]{VKK21}
{Vashishth}, V., {Karak}, B.~B., \& {Kitchatinov}, L. 2021, {Subcritical dynamo
  and hysteresis in a Babcock-Leighton type kinematic dynamo model}.
\newblock {\em Research in Astronomy and Astrophysics}, 21(10), 266.

\bibitem[{Vashishth} et~al., 2023]{VKK23}
{Vashishth}, V., {Karak}, B.~B., \& {Kitchatinov}, L. 2023, {Dynamo modelling
  for cycle variability and occurrence of grand minima in Sun-like stars:
  rotation rate dependence}.
\newblock {\em \mnras}, 522(2), 2601--2610.

\bibitem[{Waldmeier}, 1935]{W35}
{Waldmeier}, M. 1935, {Neue Eigenschaften der Sonnenfleckenkurve}.
\newblock {\em Astronomische Mitteilungen der Eidgen{\"o}ssischen Sternwarte
  Zurich}, 14, 105--136.

\bibitem[{Waldmeier}, 1955]{W55}
{Waldmeier}, M. 1955, {Ergebnisse und Probleme der Sonnenforschung.}
\newblock {\em Ergebnisse und Probleme der Sonnenforschung (Leipzig: Geest \&
  Portig)},.

\end{thebibliography}

\end{document}